\documentclass[referee]{raa}            
\pdfoutput=1 
\usepackage{graphicx,times}             
\usepackage{natbib}
\usepackage{amssymb,amsmath}
\usepackage{xcolor} 
\bibpunct{(}{)}{;}{a}{}{,}

\def\withmark{0}
\if\withmark1
    \newcommand\delq[1]{{\color{red} #1}}
    \newcommand\delu[1]{{\color{red}\sout{#1}}}
    \newcommand\addu[1]{{\color{blue} #1}}
    \hypersetup{colorlinks = true, linkcolor = black, anchorcolor = black, citecolor = black, filecolor = black, pagecolor = black, urlcolor = black}
\else 
\if\withmark0
    \newcommand\delq[1]{}
    \newcommand\delu[1]{}
    \newcommand\addu[1]{#1}
    \newcommand\addv[1]{#1}
    \newcommand\delv[1]{}
\else
\if\withmark2
    \newcommand\delq[1]{}
    \newcommand\delu[1]{}
    \newcommand\addu[1]{#1}
    \newcommand\delv[1]{{\color{red}\sout{#1}}}
    \newcommand\addv[1]{{\color{blue} #1}}
    \hypersetup{colorlinks = true, linkcolor = black, anchorcolor = black, citecolor = black, filecolor = black, pagecolor = black, urlcolor = black}
\else
    \newcommand\delu[1]{}
    \newcommand\addu[1]{{\color{red} #1}}
    \hypersetup{colorlinks = true, linkcolor = black, anchorcolor = black, citecolor = black, filecolor = black, pagecolor = black, urlcolor = black}
\fi\fi\fi

\begin{document}

\title{Lunar Cratering Asymmetries with High \delu{Lunar}\addu{Orbital} Obliquity and Inclination \addu{of the Moon}} 


\volnopage{Vol.0 (20xx) No.0, 000--000}      
\setcounter{page}{1}                         

\author{Huacheng Li
    \inst{1}
\and Nan Zhang
    \inst{1,2}
\and Zongyu Yue
    \inst{3}
\and Yizhuo Zhang
    \inst{1}
}

\institute{
    \addv{Key Laboratory of Orogenic Belts and Crustal Evolution,} School of Earth and Space Sciences, Peking University, 100871 Beijing, China; 
    {nan\_zhang@pku.edu.cn} \\
\and
    Earth Dynamics Research Group, School of Earth and Planetary Sciences, Curtin University, 6102 WA, Australia; \\
\and
    State Key Laboratory of Remote Sensing Science, Aerospace Information Research Institute, Chinese Academy of Sciences, 100101 Beijing, China;
    {yuezy@radi.ac.cn} \\
\vs\no
   {\small Received~~20xx month day; accepted~~20xx~~month day}
}

\abstract{
    \addu{Accurate estimation of cratering asymmetry on the Moon is crucial for understanding Moon evolution history. 
    Early studies of cratering asymmetry have omitted the contributions of high lunar obliquity and inclination. 
    Here, we include lunar obliquity and inclination as new controlling variables to derive the cratering rate spatial variation as a function of longitude and latitude. 
    With examining the influence of lunar obliquity and inclination on the asteroids population encountered by the Moon, we then have derived general formulas of the cratering rate spatial variation based on the crater scaling law.
    Our formulas with addition of lunar obliquity and inclination can reproduce the lunar cratering rate asymmetry at the current Earth-Moon distance and predict the apex/ant-apex ratio and the pole/equator ratio of this lunar cratering rate to be 1.36 and 0.87, respectively.
    The apex/ant-apex ratio is decreasing as the obliquity and inclination increasing.
    Combining with the evolution of lunar obliquity and inclination, our model shows that the apex/ant-apex ratio does not monotonically decrease with Earth-Moon distance and hence the influences of obliquity and inclination are not negligible on evolution of apex/ant-apex ratio.
    This model is generalizable to other planets and moons, especially for different spin-orbit resonances. }
}
    \authorrunning{Huacheng Li, Nan Zhang, Zongyu Yue \& Yizhuo Zhang }            
    \titlerunning{Lunar cratering asymmetries with high \delu{lunar} \addu{orbital} obliquity and inclination \addu{of the Moon}}  

\maketitle

\section{Introduction}
Cratering asymmetry on the lunar surface has been recognized in many studies \citep{LeFeuvre2011,Wang2016}.
Understanding of such asymmetry alters the basis of lunar cratering chronology \citep{Hiesinger2000,Fassett2012}, because it has assumed cratering rate is spatially uniform on the whole Moon \citep{McGill1977}, which eventually influences the fundamental understanding of lunar evolution. 
Quantifying the asymmetry can rectify the deviation in counting the lunar craters sampled by Apollo and Luna missions \citep{Hartmann1970,Neukum1975,Neukum1984}. 
Cratering asymmetry has been also generalized to the surface datings of other planets or moons \citep{Horedt1984,Neukum2001,Neukum2001b,Hartmann2001,Zahnle2001,Korycansky2005}.
Various \delu{cratering asymmetries }\addu{factors affecting the cratering asymmetry} on the Moon have been intensively investigated \citep{Hartmann1970,Neukum1975,Neukum1984,LeFeuvre2011,Wang2016}, and the key factors affecting the cratering asymmetry include (1) the speed and inclination of asteroids encountering the Moon \citep{LeFeuvre2011} and (2) the distance between the Earth and the Moon \citep{Zahnle2001,LeFeuvre2011,Wang2016}.

Three types of cratering asymmetries, i.e., the leading/trailing asymmetry, pole/equator asymmetry, and near/far-side asymmetry have been recognized \citep[e.g., ][]{LeFeuvre2011,Wang2016}.
The leading/trailing asymmetry has been explained by both theoretic derivations \citep{Horedt1984,LeFeuvre2011,Wang2016} and numerical simulations \citep{Gallant2009,Wang2016}. 
It has been confirmed that the leading surface receives more impactor fluxes and higher impact speed than the trailing surface due to the synchronous rotating, while this difference declines with the Earth-Moon distance increased \citep{LeFeuvre2011,Wang2016}. 
The pole/equator asymmetry has also been numerically modelled \citep{Gallant2009,Wang2016}, which suggested that low latitude of the Moon receives more impactor fluxes for the gathering
 of low inclination asteroids \citep{LeFeuvre2008,LeFeuvre2011}. 
In addition, the pole/equator asymmetry \delu{is not affected by the Earth-Moon distance (Le~Feuvre \& Wieczorek 2011). }\addu{is found to vary by less than \%1 when the Earth-Moon distance is between 20 and 60 Earth radii  \citep{LeFeuvre2011}.}
The mechanism of near/far-side asymmetry has not reached a consensus \citep{Wiesel1971,Bandermann1973}.
In previous studies, two factors affecting impact asymmetry, i.e., \delu{lunar }\addu{orbital} obliquity and inclination \addu{of the Moon} (relative to the ecliptic), have been usually neglected \citep{LeFeuvre2011,Wang2016}.
However, these two factors might be important 
\delu{during the early evolution stage }\addu{within the first 35 Earth radii of Earth-Moon distance}
when the Moon quickly left Earth \citep{Cuk2016,Ward1975}. 
Therefore, it is necessary to investigate the influences of these two factors \delu{to }\addu{on} the lunar cratering asymmetries.

In this study, we derive the impact asymmetry reliance on the \delu{lunar }\addu{orbital} obliquity and inclination \addu{of the Moon} 
\delu{by extending previous formulations in three dimensions (Le~Feuvre \& Wieczorek 2011; Wang \& Zhou 2016) }\addu{by improving previous empirical models of leading/trailing and pole/equator asymmetries \citep{LeFeuvre2011} and extending two-dimensional analytic formulas \citep{Wang2016} to the complete formulas based on three-dimensional geometry}. 
\addu{\cite{LeFeuvre2011} assumed the orbital obliquity of the Moon was constant when the Earth-Moon distance is larger than 20 Earth radii. \cite{Wang2016} calculated the cratering asymmetries in a planar model which excludes the influences of the orbital obliquity and inclination of the Moon.}
Our analytical formulation including obliquity and inclination can reveal more features of lunar leading/trailing asymmetry  \citep{LeFeuvre2011} and add explicit term for the pole/equator asymmetry \citep{Wang2016}. 
\delu{In section 2, by extending previous formulations (Wang \& Zhou 2016), the key variable in our formulation, i.e., cratering rate, is estimated based on the concentration of asteroids encountering with the Moon and scaling laws that convert asteroids velocities and diameters to the diameters of craters (Holsapple \& Housen 2007). 
In Section 3, the influences of lunar obliquity and inclination are calculated. 
In Section 4, we verify our formulas by comparing with previous results. }\addu{In Section 2, we derived the formulas for the distribution of impact flux, normal speed, and cratering rate on the Moon using the concentration of asteroids encountering with the Moon and scaling laws that convert asteroids velocities and diameters to the diameters of craters \citep{Holsapple2007}. 
Section 3 shows the resultant distributions of impact flux, normal speed, and cratering rate based on formulas in Section 2.
This result section also estimates the evolution of the apex/ant-apex ratio of cratering rate according to the evolution of orbital obliquity and inclination with different Earth-Moon distances \citep{Cuk2016}.
In Section 4, we verify formulas in Section 2 by comparing with previous results and explain how the orbital obliquity and inclination influence the lunar cratering rate asymmetry.}
Additionally, the influences of \delu{lunar }\addu{orbital} obliquity and inclination \addu{of the Moon} on the concentration of asteroids encountering with the Moon are detailed in the appendix.

\section{Method}
This section shows how we calculate the distribution of asteroids impact flux, impact speed and cratering rate using variables in Table 1.
Section 2.1 introduces assumptions and coordinate systems with which we derive the expression of asteroid's velocity $\vec{v_p}$ and the normal vector $\vec{n}$ at the impact site.
$\vec{v_p}$ and $\vec{n}$ will be used in the following calculations.
Section 2.2 uses equations from \cite{Wang2016} and \cite{LeFeuvre2011} to estimate the impact flux at different impact sites. 
These equations are rewritten as functions of $\vec{v_p}$ and $\vec{n}$.
Section 2.3 calculates the cratering rate variation using scaling law from \cite{Holsapple2007}.
Obtaining the cratering rate variation requires the impact flux variation and impact normal speed variation. 
The former has been calculated in section 2.2 and the later can be calculated with minor changes in calculation of impact flux.
\begin{table}
\begin{center}
\caption[]{ Variables or Parameters Used in the Method.}
 \begin{tabular}{lcl}
  \hline\noalign{\smallskip}
Description & Notation & Range                 \\
  \hline\noalign{\smallskip}
   inclination of asteroids' encounter velocity  & $\phi_p$    &   $[-\frac{\pi}{2},\frac{\pi}{2}]$   \\
   azimuth of asteroids' encounter velocity      & $\lambda_p$ &   $[0,2\pi]$    \\
   encounter speed of asteroids                 & $v_p$       &   $[19km/s,\sim20km/s]$ \\
   lunar orbit inclination            & $i_1$       &   $[0,\frac{\pi}{2}]$   \\
   lunar obliquity relative to the ecliptic      & $i_2$       &  $[0,\frac{\pi}{2}]$     \\
   azimuth of lunar orbit normal      & $\omega_1$  &   $[0,2\pi]$    \\
   azimuth of lunar spin axis         & $\omega_2$  &    $[0,2\pi]$   \\
   lunar true anomaly                 & $f_m$       &    $[0,2\pi]$   \\
   lunar eccentric anomaly            & $E$         &   $[0,2\pi]$    \\
   lunar mean anomaly                 & $M$         &    $[0,2\pi]$   \\
   lunar argument of perihelion       & $\omega_3$  &     $[0,2\pi]$  \\
   longitude of impact sites          & $\lambda$   &    $[0,2\pi]$   \\
   latitude of impact sites           & $\phi$      &      $[-\frac{\pi}{2},\frac{\pi}{2}]$   \\
   semi-major axis of lunar orbit     & $a_m$       &     $[25R_e,60 R_e]$ \\
   eccentricity of lunar orbit       & $e$          &     [0,1)  \\              
  \noalign{\smallskip}\hline
\end{tabular}
\end{center}
\end{table}

\subsection{Asteroids Velocity and Normal Vector at Impact Site}
\delu{We introduce the model assumptions and coordinate systems including the Moon, Earth, and asteroids.
The velocity and normal vector based on those assumption and coordinate are then solved accordingly.}
\newcommand{\rmn}[2][\omega_3]{R_z(\frac{\pi}{2} + \omega_1)R_x(i_1)R_z(#1)\vec{#2}}

This model assumes the orbit of the Moon is an ellipse with the Earth as a focus. Then in the \addu{geocentric} ecliptic coordinate system \addu{(Z-axis is parallel to the ecliptic normal and X-axis is towards mean equinox of the J2000 epoch)}, the position and velocity of the Moon are $\vec{r_m}$ and $\vec{v_m}$. 
\addv{We note that the influence of variation of $i_1,\omega_1$ or $\omega_3$ on the lunar velocity can be estimated using Eqs. (2-5) of \cite{Cuk2004} and is $<1\%$
compared to the influence of variation of $f_m$. We hence ignored the variation of $i_1,\omega_1$ or $\omega_3$ in deriving the lunar velocity.} In Eq. (2), $G$ and $M_e$ are the gravitational constant and the mass of the Earth respectively.
\begin{align}
   \vec{r_m} & = \rmn{r}  \\
   \addu{\vec{r}}   & = 
   \addu{
        \left[
        \cfrac{a_m(1-e^2)}{1+e\cos{f_m}}\cos{f_m},
        \cfrac{a_m(1-e^2)}{1+e\cos{f_m}}\sin{f_m},0
        \right]^T} \nonumber \\
   \vec{v_m} &= \rmn{v} \\
   \addu{\vec{v}}   &= 
        \addu{\left[
        -\sqrt{\frac{GM_e}{a_m(1-e^2)}}\sin{f_m},
        \sqrt{\frac{GM_e}{a_m(1-e^2)}}(e + \cos{f_m}),
        0
        \right]^T} \nonumber  \\
    & R_x(\theta) = \left[
        \begin{array}{ccc}
    1 & 0 & 0 \\
    0 & \cos{\theta} & -\sin{\theta} \\
    0 & \sin{\theta} & \cos{\theta} \\
        \end{array}
    \right], \ 
    R_z(\theta) = \left[
        \begin{array}{ccc}
    \cos{\theta} & -\sin{\theta} & 0\\
    \sin{\theta} & \cos{\theta}  & 0\\
    0 & 0 & 1
        \end{array}
    \right] \nonumber
\end{align}

\delv{The asteroids encounter with the Moon at a population concentration \addu{$C_0$} }\addv{The population concentration $C_0$ of asteroids encountering with the Moon is defined as the distribution of the relative number of asteroids that encounter with the Moon within a unit time and it can be} determined by their velocities \addu{\citep[e.g., Figure 5 of ][]{LeFeuvre2008}}. 
\delu{In our calculations, we assume this concentration $C_0$ is a function of inclination and azimuth of asteroids velocity.} 
In Eq. (3) and (4), $\vec{v_p}$ and $\vec{e_z}$ are the asteroids' encounter velocity in the \addu{geocentric} ecliptic coordinate system and an unit vector parallel to the positive Z-axis respectively.
$v_p$ is the average encounter velocity of the asteroids \addu{related to the Earth}. \delu{It is usually $19 \sim 20 km/s$ (Zahnle {et~al.} 2001; Gallant {et~al.} 2009; Le~Feuvre \& Wieczorek 2008).}
\begin{align}
    C_0 &= p(\vec{v_p}) = p(\lambda_p,\phi_p) \\
    \vec{v_p} &= v_p
   R_z(\frac{\pi}{2} + \lambda_p)R_x(\frac{\pi}{2}-\phi_p)\vec{e_z}   
\end{align}
\delu{where the marginal distribution of Eq. (3) $\int_{-\pi}^{\pi} C_0 d\lambda_p$ is taken from Le~Feuvre \& Wieczorek(2011) and Bottke {et~al.}(2002).}
\addu{In this model, $\vec{v}_p$ is determined by $\lambda_p$, $\phi_p$, and $v_p$.}
\delu{In this study} The concentration of asteroids encountering with the Moon is \delu{thought }\addu{assumed} unaffected by the orbital obliquity and inclination of the Moon (see the appendix \addu{A}). 
\addu{Eq. (A.20) indicates the concentration of asteroids encountering with the Moon can be estimated by the concentration encountering with the Earth. 
Then $C_0$ should be function of $(v_p,\lambda_p,\phi_p)$.
Spectrum of $v_p$ is not considered in this study and it is set as the average encounter speed \citep{Horedt1984, Zahnle2001}.
Then $C_0$ is independent of $v_p$ and a function of $(\lambda_p,\phi_p)$.
Since the precession of lunar orbit, the asteroids' azimuth distribution will not affect the cratering asymmetries.
Therefore, only the marginal distribution of Eq. (3) $\int_{-\pi}^{\pi} C_0 d\lambda_p$ is required in the calculation of cratering asymmetries. 
This marginal distribution is taken from \cite{LeFeuvre2008} and it has been shown in Figure 6 of \cite{LeFeuvre2008}.}

\begin{figure}
\centering\includegraphics[width=0.7\textwidth, angle=0]{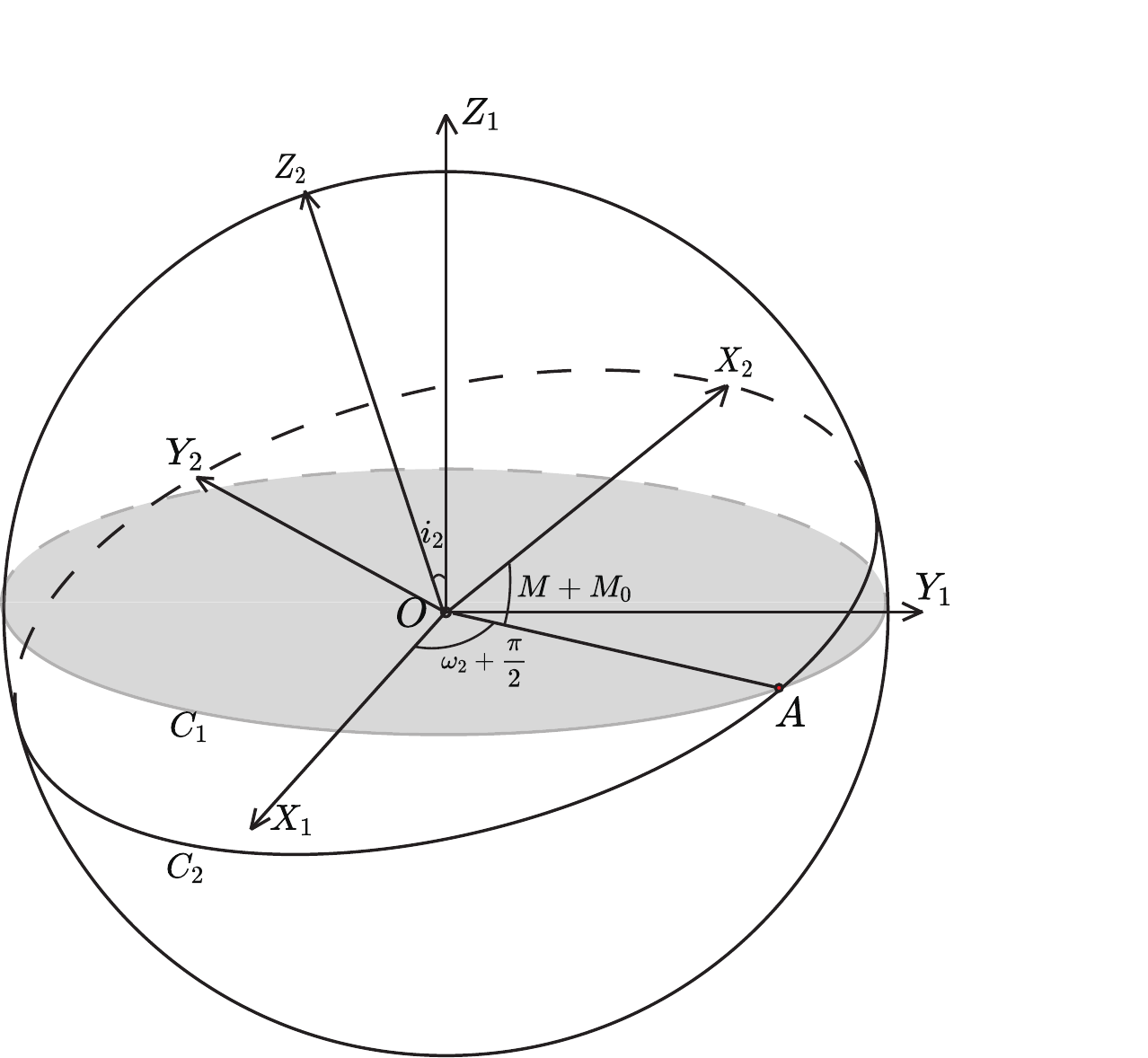}
\caption{The coordinate systems used in calculation. The geocentric ecliptic coordinate system is $OX_1Y_1Z_1$. The lunar fixed coordinate system is $OX_2Y_2Z_2$ and its origin is translated to the Earth. The gray plane $C_1$ is the ecliptic plane. The plane $C_2$ is the lunar equatorial plane. $\overline{OA}$ is the intersection of $C_1$ and $C_2$.}
\end{figure}

The Moon is \addv{assumed to be} synchronously rotating with a constant angular velocity and the prime meridian is determined by the mean sub-Earth point \citep{LRO2008}.
In the lunar fixed coordinate system whose X-axis is the intersection of the lunar equator plane and prime meridian plane, and Z-axis is the lunar spin axis, the normal at lunar surface is $\vec{n}(\lambda,\phi)$ and the transformation matrix $T$ \delu{the ecliptic coordinate system and this lunar fixed coordinate system}
\addu{from this lunar fixed coordinate system to the geocentric ecliptic coordinate system}
 are determined by $\omega_2$, $i_2$ and $M$.
 \addu{The relationship between coordinate systems used in this section is illustrated in Figure 1.}
\newcommand{\Tm}[1][M+M_0]{R_z(\frac{\pi}{2} + \omega_2)R_x(i_2)R_z(#1)}
\newcommand{\nn}{R_z(\frac{\pi}{2} + \lambda)R_x(\frac{\pi}{2}-\phi)}
\begin{align}
    \vec{n}(\lambda,\phi) &= \nn \vec{e_z} \\
    T &= \Tm
\end{align}

\addv{In Eq. (6), $M_0$ is a parameter related to the position of the mean sub-Earth point and it will be determined by Eqs. (7-9).}
When the Moon at the perigee, the center of Earth passes through the lunar prime meridian plane.
\begin{align}
    & \frac{\vec{r_m}}{|\vec{r_m}|} \Big{|}_{f_m =0}+ T \cdot \vec{n}(0,\phi) = 0 \label{equM0}
\end{align}
Solving Eq. (\ref{equM0}), this gives 
\newcommand{\MA}{
     \cos{i_1}\sin{(\omega_1-\omega_2)}\sin{\omega_3}
     - \cos{(\omega_1-\omega_2)}\cos{\omega_3}
}
\newcommand{\MB}{
    - \cos{i_1}\cos{i_2}\cos{(\omega_1-\omega_2)}\sin{\omega_3}
    - \cos{i_2}\sin{(\omega_1-\omega_2)}\cos{\omega_3}
    + \sin{i_1}\sin{i_2}\sin{\omega_3}
}
\newcommand{\MAB}{
    \sin{i_2}\sin{(\omega_1-\omega_2)}\cos{\omega_3}
    - \cos{i_1}\sin{i_2}\cos{(\omega_1-\omega_2)}\sin{\omega_3}
    - \sin{i_1}\cos{i_2}\sin{\omega_3}
}
\begin{align}
	\cos{M_0} &= \frac{\MA}{\sqrt{1- (\MAB)^2}} \\
	\sin{M_0} &= \frac{\MB}{\sqrt{1- (\MAB)^2}} 
\end{align}

\subsection{Distribution of Impact Flux}
For given $\vec{v_p}$, $\vec{r_m}$ and $\vec{v_m}$, the velocity of asteroids relative to the Moon is $\vec{v_p} - \vec{v_m}$. 
Define $\hat{x}$ as $\-\cfrac{\vec{v_p} - \vec{v_m}}{|\vec{v_p} - \vec{v_m}|} \cdot (T\vec{n}(\lambda,\phi))$. 
\addv{The impact flux $\delta F$ is defined as the distribution of the number of asteroids that impact on the lunar surface within a unit area and a unit time.} 
According to Eq. (26) of \cite{Wang2016} and Eq. (A.47) of \cite{LeFeuvre2011}, the impact flux $\delta F$ is a function of $\hat{x}$.
In Eq. (13), $M_m$ and $R_m$ are the mass and radius of the Moon respectively.
\begin{align}
    \delta F &= C_0 |\vec{v_p} - \vec{v_m}| f(\hat{x}) \\
    f_1(\hat{x}) &= \left\{\begin{array}{lcl}
    \hat{x} & , & \hat{x} \ge 0 \\
    0 & , & \hat{x} < 0
    \end{array}\right. \\
    f_2(x) &= \left\{\begin{array}{lcl}
    \frac{1}{4}(1+\Gamma)^{-1}(1+\mu^{-1})(\Gamma + (1+\mu)\hat{x})& , & \hat{x} \ge \frac{-\Gamma}{2+\Gamma} \\
    0 & , & \hat{x} < \frac{-\Gamma}{2+\Gamma}
\end{array} \label{flux}
    \right.
    \\
    \mu = &\sqrt{1 + \cfrac{2\Gamma}{1+\hat{x}}}
    \ ,\Gamma = \cfrac{2GM_m}{|\vec{v_p} - \vec{v_m}|^2 R_m}
\end{align}
where $f_1(\hat{x})$ and $f_2(\hat{x})$ are two forms of $f(\hat{x})$. $f_1(\hat{x})$ is from  \cite{Wang2016} which assumes the trajectories of asteroids are straight lines in the direction of their common encounter velocity. 
While $f_2(\hat{x})$ is from  \cite{LeFeuvre2011} in which trajectories of asteroids are treated as hyperbolic curve with a focus at the center of the Moon. 
Because $\Gamma < 0.02$, we can expand Eq. (\ref{flux}) around 0 with Taylor series.
\begin{align}
f_2(\hat{x}) = \hat{x} + (\frac{1}{2} - 2\hat{x})\Gamma 
+ \frac{4\hat{x}^3 + 6\hat{x}^2 -3}{4(1+\hat{x})^2}\Gamma ^2 + o(\Gamma^3) \ \ \hat{x} > \frac{-\Gamma}{2+\Gamma}
\end{align}
Obviously, $f_1(\hat{x})$ is the first order approximation of $f_2(\hat{x})$. 
The absolute relative difference between $f_1$ and $f_2$ is less than $\%3.5$.
For simplicity, we use $f(\hat{x}) = f_1(\hat{x})$ in following calculations. 
Then the average flux within a period of lunar orbit is 
\begin{align}
    &F = \frac{1}{2\pi}\int_{0}^{2\pi}  \delta F dM 
    = \frac{1}{2\pi}\int_{0}^{2\pi} C_0 |\vec{v_p} - \vec{v_m}| f(\hat{x}) dM 
\end{align} 
It is known today that $\omega_3$ changes with a period of 8.85 years and $\omega_2$ changes with a period of 18.61 years. 
The secular average flux independent of them is 
\begin{align}
    &\overline{F}(\lambda,\phi;a_m,e,i_1,i_2,v_p) = \int_{0}^{2\pi}\frac{d\omega_3}{2\pi}
\int_{0}^{2\pi}\frac{d\omega_2}{2\pi}
\int_{0}^{2\pi}\frac{d\lambda_p}{2\pi}
\int_{-\frac{\pi}{2}}^{\frac{\pi}{2}}
F\frac{d\phi_p}{\pi}
\end{align} 

\subsection{Distribution of Normal Impact Speed and Cratering Rate}
In this model, the impact angle of asteroids at lunar surface is $\theta$ \citep[Eq. (A.54) of][]{LeFeuvre2011} and the normal impact speed \delv{can be expressed as }\addv{is $V_{\perp}$
\citep[Eqs. (A.50-A.51) of][]{LeFeuvre2011}.}
\begin{align}
    V_{\perp} &= |\vec{v_p} - \vec{v_m}| \sqrt{1 + \Gamma} \sin{\theta}
    = |\vec{v_p} - \vec{v_m}| g(\hat{x})
\end{align}
$g(\hat{x})$ can also be written as two different forms $g_1(\hat{x})$ and $g_2(\hat{x})$. $g_1(\hat{x})$ is from \cite{Wang2016} and $g_2(\hat{x})$ is from  \cite{LeFeuvre2011}.
\begin{align}
    g_1(\hat{x}) &= \hat{x}/\sqrt{1 + \Gamma} \\
    g_2(\hat{x}) &= \sqrt{1 + \Gamma - (\frac{1+\mu}{2})^2(1-\hat{x}^2)} \nonumber \\
    & = |\hat{x}| + \frac{1}{2} sgn(\hat{x}) \Gamma - \frac{sgn(\hat{x})}{4(1+\hat{x})} \Gamma^2 + o(\Gamma^3) \label{velc}
\end{align}
Similar to $f(\hat{x})$, we expand $g_2(\hat{x})$ around $\Gamma = 0$. 
$g_1(\hat{x})$ is the first order approximation of $g_2(\hat{x})$. Substituting $g_1(\hat{x})$ in to Eq. (17). 
We obtain the average normal speed  
\begin{align}
& \overline{V}_{\perp}(\lambda,\phi;a_m,e,i_1,i_2,v_p) = \frac{1}{\overline{F}(\lambda,\phi;a_m,e,i_1,i_2,v_p)} \nonumber \\
& \cdot \int_{0}^{2\pi}\frac{d\omega_3}{2\pi}
\int_{0}^{2\pi}\frac{d\omega_2}{2\pi}
\int_{0}^{2\pi}\frac{d\lambda_p}{2\pi}
\int_{-\frac{\pi}{2}}^{\frac{\pi}{2}}\frac{d\phi_p}{\pi}
\int_{0}^{2\pi}
    \delta F V_{\perp}
\frac{dM}{2\pi}
\end{align}
Combing Eq. (16) and Eq. (20), we finally obtain the cratering rate expression.
Similar to Eq. (56) of \cite{Wang2016} and applying the scaling law of crater diameters(e.g., \cite{Holsapple2007}), the cratering rate in our model is  
\begin{align}
    & N_c(\lambda,\phi;a_m,e,i_1,i_2,v_p) 
     \propto (\overline{V}_{\perp}(\lambda,\phi;a_m,e,i_1,i_2,v_p))^{\gamma_p \alpha_p} \overline{F}(\lambda,\phi;a_m,e,i_1,i_2,v_p)
\end{align}
Here the cratering rate calculation only takes account the the near-Earth objects: $\gamma_p \alpha_p = 0.987$ \citep{Bottke2002,Holsapple2007}.
$\alpha_p$ is an exponent in the cumulative size distribution of near-Earth objects diameter \citep{Bottke2002}.
$\gamma_p$ is a parameter in the scaling law \citep{Holsapple2007,LeFeuvre2011,Wang2016}.

\section{Result}
In this section, we describe the cratering rate asymmetry produced by Eq. (21). 
Section 3.1 demonstrates the \delu{cratering rate} spatial variation of \addu{impact flux, normal speed, and cratering rate}. \delu{with parameters set at current values of the Earth-Moon system.}
Section 3.2 reveals the influences of orbital obliquity and inclination of the Moon on lunar cratering rate asymmetry. 
Section 3.3 provides the evolution of apex/ant-apex ratio with orbital obliquity and inclination of the Moon. 
\subsection{Spatial Variations of \addu{Impact Flux, Normal Speed, and} Cratering Rate}
\addu{First, our derived formula is used to calculate the cratering rate spatial variation at current values of the Earth-Moon system since such variation can be compared with previous predictions by \cite{LeFeuvre2011} and \cite{Wang2016}.}
Figure 2 shows the relative spatial variation of impact flux, normal speed, and cratering rate on the Moon \addu{with parameters set at current values of the Earth-Moon system}. 
The parameters involved in Eq. (21) are set as $(a_m,e,i_1,i_2,v_p) = (60R_e,0.0549,5.145^\circ,1.535^\circ,19km/s)$ ($R_e$ is the radius of the Earth). 
In Figure 2, the relative cratering rate is symmetry about $0^\circ N$. This symmetry arises from the symmetry of the asteroids's concentration $C_0$.
The maximum of \delu{cratering rate} \addu{impact flux} occurs at $(90^\circ W,0^\circ N)$ and the minimum is at $(90^\circ E,\pm 65^\circ N)$. The maximum/minimum ratio \delu{is 1.40} \addu{of impact flux is 1.24}.
\addu{The maximum of normal speed occurs at $(90^\circ W,0^\circ N)$ and the minimum is at $(90^\circ E,\pm 47^\circ N)$. The maximum normal speed is 13.7 $km/s$ and the minimum is 12.1 $km/s$.
The maximum of cratering rate occurs at $(90^\circ W,0^\circ N)$ and the minimum is at $(90^\circ E,\pm 53^\circ N)$. The maximum/minimum cratering rate ratio is 1.40.}
The apex/ant-apex ratio (the cratering rate ratio between $(90^\circ W,0^\circ N)$ and $(90^\circ E,0^\circ N)$) is 1.36 and this ratio is a measure of the longitudinal variation. 
The pole/equator ratio is 0.87 and this is a measure of the latitudinal variation.
\addu{The impact flux, normal speed, and cratering rate with $e =0$ (other parameters are set same as Figure 2) are also calculated.
The relative difference of cratering rate ($e = 0$) from Figure 2 is less than 0.2\%.}
\begin{figure}
\centering\includegraphics[width=0.95\textwidth,angle=0]{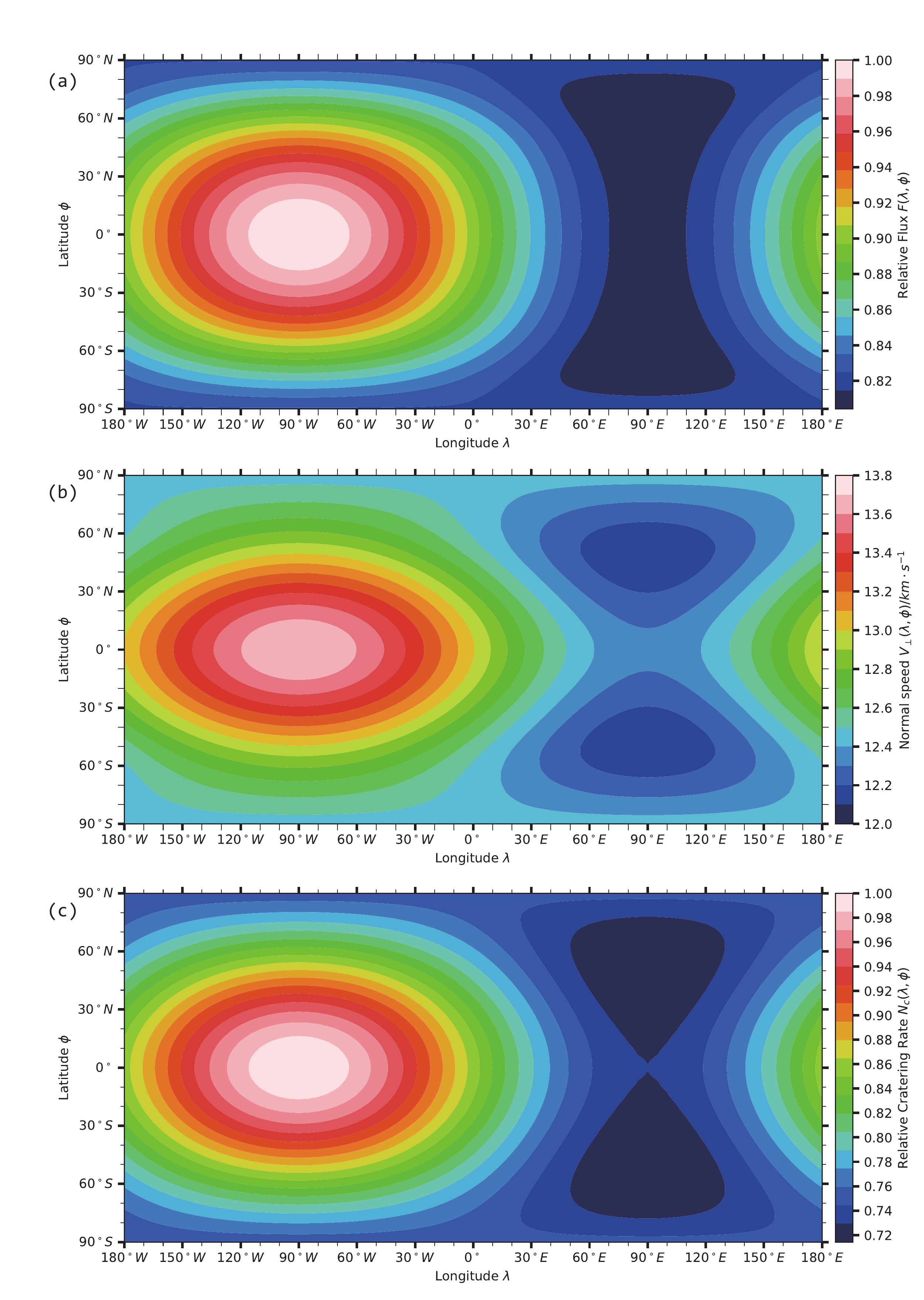}
\caption{Distribution of impact flux (a), normal speed (b), and cratering rate (c) on the Moon for the current lunar orbital obliquity, inclination and Earth-Moon distance. The maximum is set to 1.00.}
\end{figure}

\subsection{Influences of Orbital Inclination and Obliquity of the Moon}
\addu{We next investigate the specific effects of lunar orbital inclination and obliquity on apex/ant-apex and pole/equator ratios.}
Figure 3 shows the apex/ant-apex ratio $r_1$ and pole/equator ratio $r_2$  with different lunar inclination and obliquity. $i_2$ is not same with the lunar obliquity to its orbit normal. 
For Cassini state 2 ($\omega_1 = \omega_2 + \pi$), lunar obliquity \addv{relative to the lunar orbit normal} is $i_1 + i_2$, while for Cassini state 1 ($\omega_1 = \omega_2$), lunar obliquity \addv{relative to the lunar orbit normal} is $|i_1 - i_2|$ \citep{Ward1975}.
Other parameters in Eq. (21) are set as $(a_m,e,v_p) = (60R_e,0.0549,19km/s)$.
$r_1$ decreases with both $i_1$ and $i_2$ increased, while $r_2$ increases with increase in $i_2$ and seems to be independent of $i_1$. 
\addu{Based on our calculated $i_1$ and $i_2$ distribution, we speculate the correlation of $r_1$ or $r_2$ with $\cos{(i_1 + i_2)}$ and $\cos{(2 i_2)}$ can be fitted as linear regressions.}
\delu{The correlation of $r_1$ or $r_2$ with $i_1$ and $i_2$ can be fitted as Eq. (22) and (23).}
\begin{align}
    r_1 &= a_{11} + a_{12} \cos{(i_1 + i_2)} + a_{13} \cos{(2 i_2)} 
    \\
    r_2 &= a_{21} + a_{22} \cos{(i_1 + i_2)} + a_{23} \cos{(2 i_2)}
\end{align}
When $a_m=60R_e$, the fitting result is 
\begin{align}
\left(
\begin{matrix}
    a_{11} & a_{12} & a_{13} \\
    a_{21} & a_{22} & a_{23}
\end{matrix}
\right)
=
\left(
\begin{matrix}
    1.12676 &  0.2469790 &  0.0089461 \\
    0.97137 & -0.0029778 & -0.0930123 
\end{matrix}
\right)
\end{align}
\addu{When $i_1$ and $i_2$ are between $0^\circ$ and $45^\circ$, the relative error between fitting result and Figure 3 is less than 2.4\% for $r_1$ and 0.15\% for $r_2$.}
\begin{figure}
\centering\includegraphics[width=1.0\textwidth, angle=0]{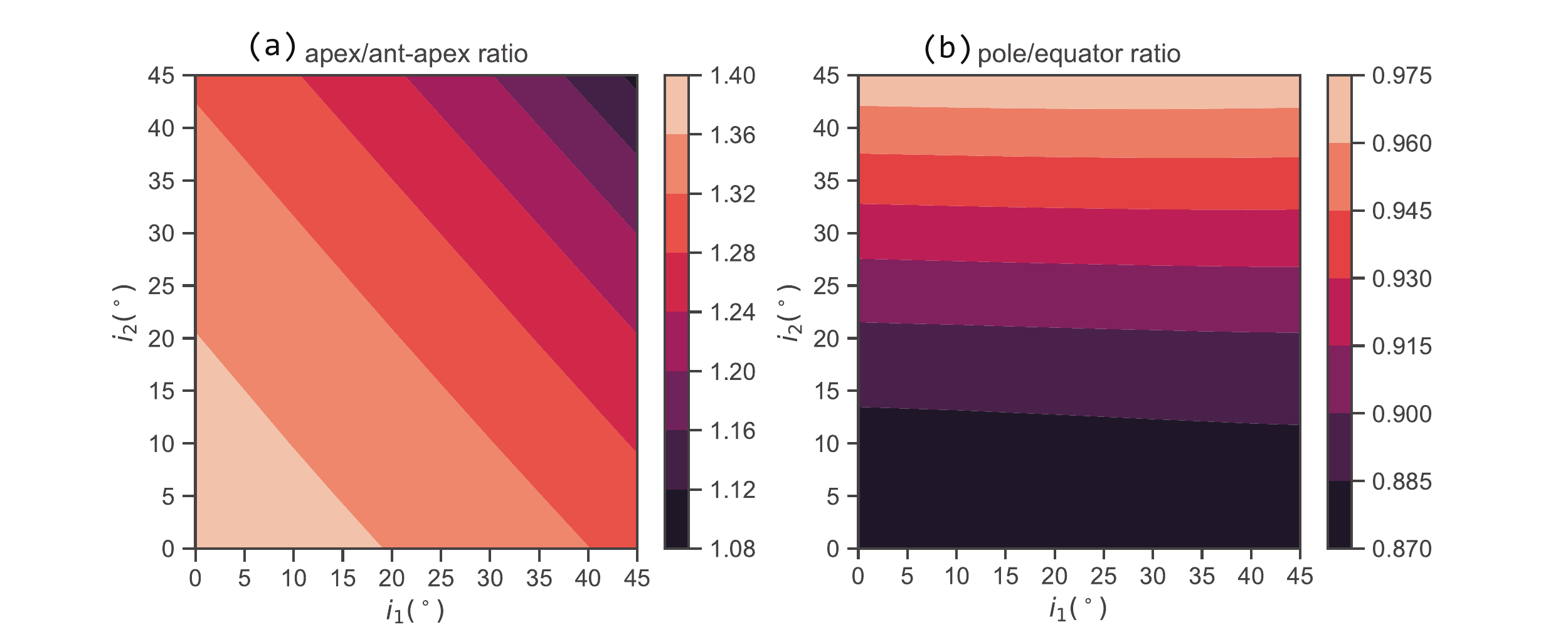}
\caption{The apex/ant-apex (a) and pole/equator (b) ratios with orbital obliquity and inclination of the Moon. Only the Cassini state 2 ($\omega_1 = \omega_2 + \pi$) is calculated.}
\end{figure}
\subsection{Evolution of the Apex/ant-apex Ratio}
The past obliquity has been very high and the lunar inclination is also different from current value \citep{Ward1975,Cuk2016}.
\delu{Through the semi-analytical model of the lunar orbital evolution from Ćuk {et~al.}(2016), we obtain the evolution of the orbital obliquity and inclination of the Moon with the Earth-Moon distance which is applied in our model to estimate the evolution of apex/ant-apex ratio (Figure 3).}
\addu{We obtain the evolution of the lunar orbital obliquity and inclination with the Earth-Moon distance by reproducing the semi-analytical method for the lunar orbital evolution from \cite{Cuk2016}. 
This method includes solving the differential equations of lunar synchronous orbit controlled by Earth and Moon tidal dissipation, as well as coupling them with the equation to satisfy the Cassini state \cite{Ward1975}. 
The solutions show that the lunar inclination damps from the initial high value to its present low value $5.1^\circ$ due to tidal dissipation, and the lunar obliquity first increases and then decreases to current value $1.5^\circ$ with the jump between 29.7$R_e$ and 35$R_e$ due to the transitions from Cassini state 1 to Cassini state 2, which is similar to the extended data Figure 1 in \cite{Cuk2016}. 
We next apply this evolution in our model to estimate the evolution of apex/ant-apex ratio (Figure 4).}
According to \cite{Cuk2016}, the Moon is in non-synchronous rotation from 29.7$R_e$ to about 35$R_e$ (gray box in Figure 4). 
When the Moon is at Cassini state 1 (the Earth-Moon distance $< 29.7R_e$), the apex/ant-apex ratio decreases with $a_m$. 
When the Moon is at Cassini state 2 ($> 35 R_e$), this ratio reaches a maximum between 40$R_e$ and 45$R_e$. 
\begin{figure}
\centering\includegraphics[width=0.8\textwidth, angle=0]{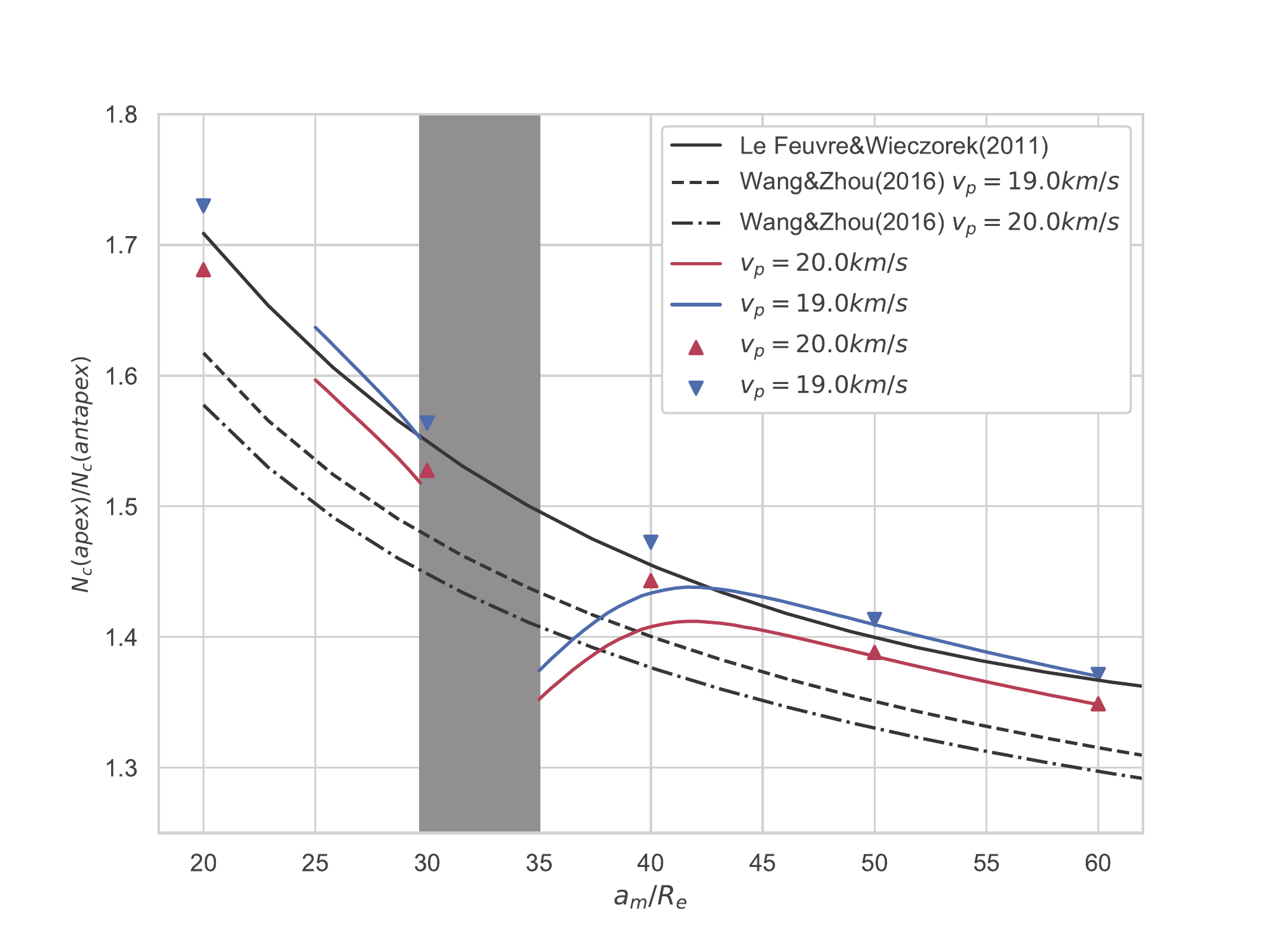}
\caption{Evolution of the apex/ant-apex ratio with Earth-Moon distance.
The X-axis represents the Earth-Moon distance and Y-axis represents the apex/ant-apex ratio.
The black solid line is $1.12e^{-0.0529a_m/R_e} + 1.32$ from  \cite{LeFeuvre2011}. The value of $\alpha_p\gamma_p$ in \cite{LeFeuvre2011} ranges from $0.907$ to $1.25$.
The other two dashed black curves represent Eq. (124) of \cite{Wang2016} with $\alpha_p\gamma_p = 0.987$.
The red and blue triangles represent results from Eq. (21) which uses constant obliquity and inclination same as current values $(i_1,i_2) = (5.145^\circ,1.535^\circ)$.
The red and blue lines is calculated basing on the evolution of orbital obliquity and inclination of the Moon from  \cite{Cuk2016} with lunar tidal dissipation number $Q_M = 38$.}
\end{figure}

\addu{
\subsection{Cratering Rate Distribution of 3:2 Resonance}
Our formulas can also predict cratering rate distributions with various spin-orbit resonance.
When the resonance is 3:2 \citep[applicable to Mercury, ][]{Colombo1965}, we have a different transformation matrix in Eq. (6) as 
\begin{align}
    T^{'} &=  R_z(\frac{\pi}{2} + \omega_2)R_x(i_2)R_z(\frac{3}{2}M + M_0)
\end{align}
Also because this resonance is 3:2, a full integration interval for Eqs. (16), (20), and (21) is extended to two periods  $(0,4\pi]$.
If setting up the parameters involved in Eq. (21) as those for Mercury $(\sqrt{GM_e/a_m},e,i_1,i_2,v_p) = (48.0km/s,0.205,7.0^\circ,7.0^\circ,42.2km/s)$.
Further substituting T in Eqs. (7-21) with T' and using the asteroids inclination distribution from \cite{LeFeuvre2008}, 
the maximum and minimum of cratering rate with 3:2 resonance is at $(\pm90^\circ E,0^\circ N)$ and $\pm90^\circ N$, respectively.
The maximum/minimum cratering rate ratio is 3.64. 
When the orbital eccentric is degraded to 0.0 \citep{LeFeuvre2008,Wang2016}, the maximum and minimum are at $0^\circ N$ and $\pm90^\circ N$, respectively. 
The maximum/minimum cratering rate ratio is 2.91.}


\section{Discussion}
\subsection{Comparison with Previous Results}
\addu{In Figure 2(c),} this study gives a similar current cratering rate spatial variation ($a_m = 60R_e$) as \cite{LeFeuvre2011} in which the maximum and minimum appear at $(90^\circ W,0^\circ N)$ and $(90^\circ E,\pm 65^\circ N)$ respectively. \addv{The difference in the location of the minimum between our result and \cite{LeFeuvre2011} may be brought by $f_1$ and $g_1$ used in our calculations.}
\addu{The apex/ant-apex ratio for current Moon from \cite{LeFeuvre2011} is 1.37.
The apex/ant-apex ratio for current Moon from \cite{Wang2016} with $v_p=19 km/s,\alpha_p\gamma_p=0.987$ and this study are 1.32 and 1.36 respectively.}
As an extension \delu{of} \addu{based on} \cite{LeFeuvre2011} and \cite{Wang2016}, this study gives a value between them.
The larger relative difference between this study and \cite{Wang2016} is \addu{probably} caused by \delu{this study considers }\addu{either } the asteroids inclination \addu{or the orbital obliquity and inclination of the Moon}. 
The pole/equator ratio \addu{in Figure 2(c)} is 0.87\delu{ in this study}.
This value is higher than 0.80 in \cite{LeFeuvre2011}.
This difference \delu{is} \addu{may be} brought by $f_1$ and $g_1$ used in our calculations. 
Besides the current cratering rate, this study also gives the evolution of the apex/ant-apex ratio in Figure 4.
The results from \cite{LeFeuvre2011} and \cite{Wang2016} are also included in Figure 4.
\addu{The value of $\alpha_p\gamma_p$ adopted in \cite{LeFeuvre2011} is about $0.907\sim 1.25$, because they used different parameters in crater scaling law (in non-porous gravity scaling regime $\gamma_p=0.564$ while in porous $\gamma_p=0.410$) and a 10th-order polynomial to fit the size distribution of asteroids ($\alpha_p \approx 2.22$).
The influences of $\alpha_p\gamma_p$ is shown in Figure 5.
When $\alpha_p\gamma_p$ is between $0.907\sim 1.25$ and $v_p$ is between $19\sim 20km/s$, the apex/ant-apex ratio for current Moon calculated by this model is about $1.33\sim 1.41$.}
\addu{Although the value of $\alpha_p\gamma_p$ or $v_p$ in this study is different from \cite{LeFeuvre2011}}, \delu{if the lunar obliquity and inclination are ignorable}
\addu{if the orbital obliquity and inclination of the Moon is assumed constant and same with current values $(i_1,i_2) = (5.145^\circ,1.535^\circ)$}, this study will reproduce the results predicted by \cite{LeFeuvre2011} (red and blue triangle in Figure 4).

If we consider the \addu{variation of} obliquity and inclination, when the Earth-Moon distance is more than $\sim 42R_e$, this model gives a consistent value with \cite{LeFeuvre2011} and a similar trend with \cite{Wang2016}.
However, when Earth-Moon distance is between $35R_e$ and $\sim 42R_e$, this study gives an opposite trend to previous results. 
According to \cite{Cuk2016}, the orbital obliquity and inclination of the Moon decrease in this interval. 
This evolution trend of apex/ant-apex ratio can be explained by the influences of orbital obliquity and inclination of the Moon.
When Earth-Moon distance is between $29.7R_e$ and $35R_e$, the Moon is in non-synchronous rotation and the apex/ant-apex ratio will be diminished by non-synchronous rotation.
When Earth-Moon distance is less than $29.7R_e$, the apex/ant-apex ratio is calculated under the assumption: the inclination distribution of asteroids' velocity is same as current distribution. 
Although the obliquity and inclination is very high, the apex/ant-apex ratio is consistent with \cite{LeFeuvre2011}.
We note that the population of asteroids is dominated by main-belt asteroids during the late heavy bombardment and and near-Earth objects since $3.8-3.7$ Ga according to \cite{Strom2015}.
The population of near-Earth objects have been in steady state for the past $\sim 3$ Ga \citep{Bottke2002}. 
The evolution of apex/ant-apex ratio for the Earth-Moon distance $< 29.7 R_e$ may be quite different \delu{with }\addu{from that} shown in Figure 4. 
It is confirmed that the influences of orbital obliquity and inclination of the Moon are not negligible in analysing lunar cratering asymmetry.
\begin{figure}
\centering\includegraphics[width=0.7\textwidth, angle=0]{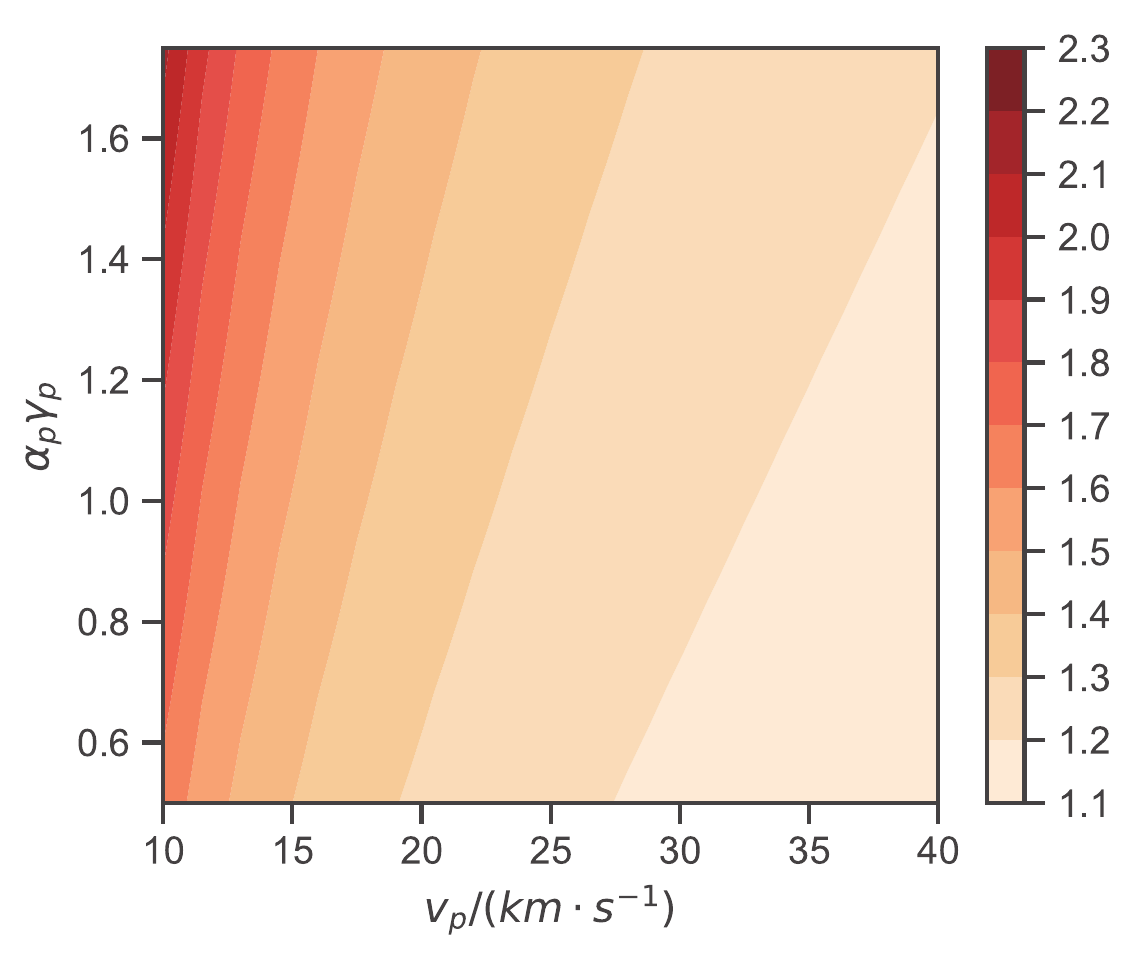}
\caption{Apex/ant-apex ratio of cratering rate with different $\alpha_p\gamma_p$ and $v_p$. Other parameters involved in Eq. (21) is set as $(a_m,e,i_1,i_2) = (60R_e,0.0549,5.145^\circ,1.535^\circ)$.}
\end{figure}

\subsection{Explanation for the Influences of Orbital Obliquity and Inclination of the Moon}
\begin{figure}
\centering\includegraphics[width=0.6\textwidth, angle=0]{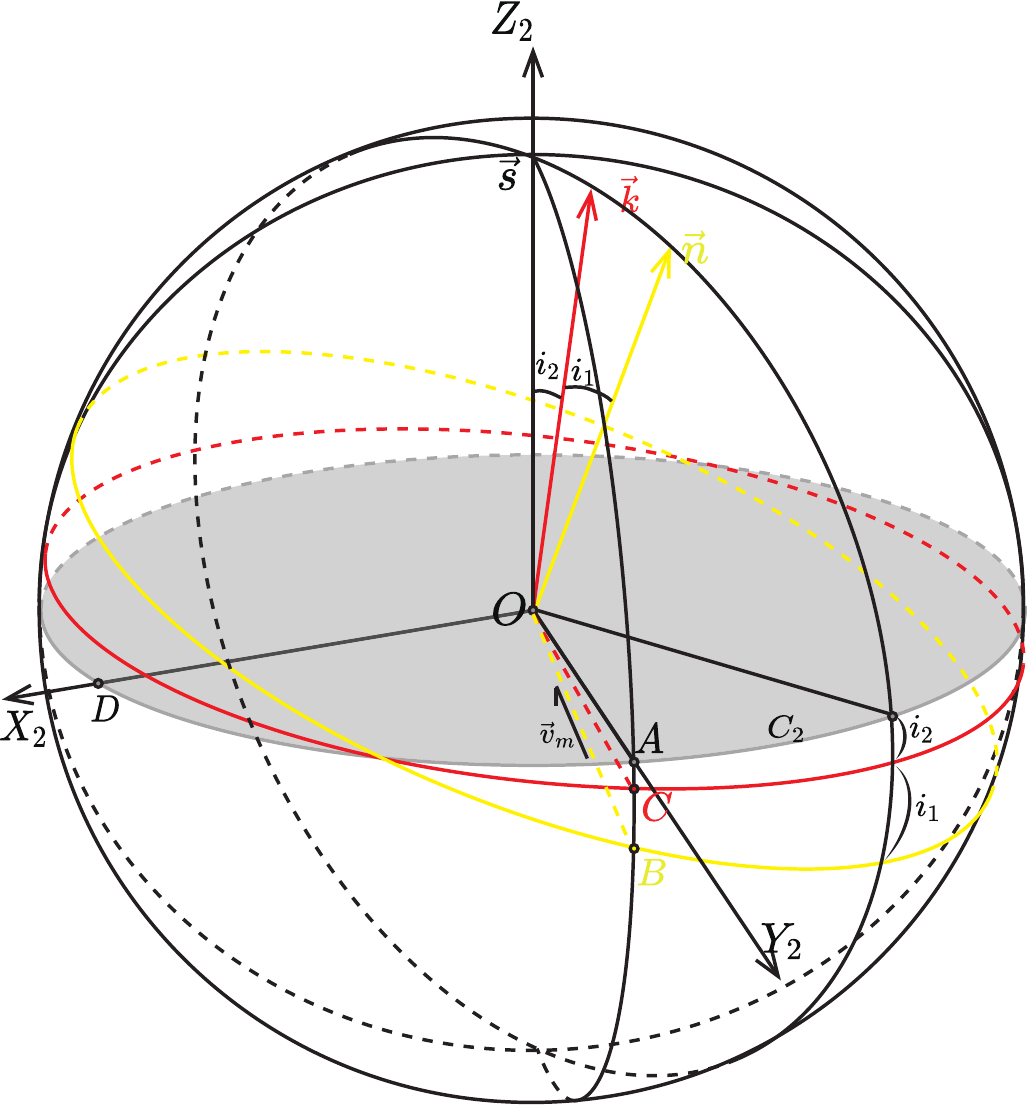}
\caption{
Sketch for our model in \delu{a Moon-centric coordinate system} \addu{the lunar fixed coordinate system $OX_2Y_2Z_2$}.
\addu{The origin $O$ is at the center of the Moon. 
The $Z_2$-axis is parallel to the lunar spin axis $\vec s$ and the $X_2$-axis points to the mean sub earth point. $\vec k$ is the ecliptic normal. $\vec{n}$ is the lunar orbit normal.}
\delu{The Z-axis is the ecliptic normal $\vec{k}$ and the X-axis points to the mean sub earth point. $\vec{s}$ is the lunar spin axis. $\vec{n}$ is the lunar orbit normal.
For a given mean anomaly $M$, $\vec{v_m}$ is the lunar velocity and the apex is the point $(-90^\circ E,0^\circ N)$. 
Point B and $A_0$ is on the lunar surface. $\overline{OB}$ is parallel to $\vec{v_m}$. $A_0$ is the intersection of Y-axis and the lunar surface.}
\addu{Point A, B, and C are on the intersection of the lunar surface and the plane $OY_2Z_2$. 
$\overline{OB}$ is parallel to $\vec{v_m}$ and perpendicular to $\vec n$.
Point A is on the lunar equator plane $C_2$ and it is the ant-apex point $(90^\circ E,0^\circ N)$.
Point C is on the ecliptic plane. Point D is the mean sub-earth point.}
}
\end{figure}
\addu{When the lunar orbit eccentric $e=0.0$}, our model is sketched in Figure 6. 
\delu{Because the coordinate system is fixed to the lunar surface and the Moon is synchronously rotating,}
\addu{In a lunar rotation period}, the relative position between $\{\vec{n},\vec{k},\vec{s}\}$ and the coordinate system \addu{$OX_2Y_2Z_2$} is not fixed.
\delu{Here we think $\{\vec{n},\vec{k},\vec{s}\}$ is fixed and the coordinate system changes with different mean anomalies.
For different mean anomalies, the locus of apex is a great circle perpendicular to $\vec{s}$ and the locus of $B$ is the great circle perpendicular to $\vec{n}$.
For the leading/trailing asymmetry which is caused by synchronously rotating, the maximum of cratering rate is at point $B$, while the leading hemisphere depends on the apex.
The farther the apex is from $B$, the smaller the leading/trailing asymmetry.}\addu{The apex or ant-apex point is on the gray cycle $C_2$.
When $i_1=i_2=0$, the gray circle $C_2$, red circle, and yellow circle coincide and $r_1$ reaches its maximum.
The influence of asteroids inclination is related to the length of $\stackrel\frown{AC}$.
The influence of lunar velocity is related to the length of $\stackrel\frown{AB}$.
}
The farther the \delu{apex} \addu{ant-apex} is from $B$ \addu{or $C$}, the smaller the leading/trailing asymmetry.
In our model, the angular distance between A \addu{(ant-apex)} and B is in $[0,max\{|i_1 + i_2|,|i_1 - i_2|\}]$ \addu{and the angular distance between A and C is in $[0,i_2]$}.
For Cassini state 2, $max\{|i_1 + i_2|,|i_1 - i_2|\} = |i_1 + i_2|$. 
This is consistent with the fitting result: \delu{apex/ant-apex ratio} \addu{$r_1$} is proportional to $\cos{(i_1 + i_2)}$ \addu{and $\cos{(2i_2)}$}.
\delu{For the pole/equator asymmetry which is caused by the distribution of asteroids' inclination to the ecliptic, its maximum is at the great circle perpendicular to $\vec{k}$ (locus of $A_0$ with different mean anomaly).
While the lunar equator is defined as the great circle perpendicular to $\vec{s}$, the angular distance between $\vec{k}$ and $\vec{s}$ is $i_2$.}
\addu{For $r_2$ which is related to the angular distance between point D and the red or yellow plane. When $i_1=i_2=0$, $r_2$ reaches its minimum. The angular distance between point D and the red circle is in  $[0,i_2]$ and 
the angular distance between point D and the yellow circle is in $[0,max\{|i_1 + i_2|,|i_1 - i_2|\}$.}
The pole/equator asymmetry will decrease with increasing in \delu{$i_2$} those two angular distances.
This is also consistent with the fitting result: $r_2$ is proportional to $\cos{(2i_2)}$ \addu{and $\cos{(i_1+i_2)}$}.

\subsection{Generalization of this Model}
The orbital obliquity and inclination of the Moon, Earth-Moon distance, lunar orbital eccentric, and lunar rotation speed have been included in this model.
In addition to the Moon, this model can be applied to other planets and moons, especially for the types of spin-orbit resonance. 
For example, Mercury is tidally locked with the Sun in a 3:2 resonance. 
\addu{The cratering rate distribution of 3:2 Resonance is detailed in the Figure 7}\delu{ Eq. (6) can be modified as}
\delq{\begin{align}
    T^{'} &=  R_z(\frac{\pi}{2} + \omega_2)R_x(i_2)R_z(\frac{3}{2}M + M_0) \nonumber
\end{align}}
\delu{Substituting $T$ in Eq. (7-21) with $T^{'}$ and using the asteroids inclination distribution from Le~Feuvre \& Wieczorek(2008), the maximum of cratering rate with 3:2 resonance is at $(\pm 60^\circ W,0^\circ N)$. 
The minimum is at $(180^\circ E,0^\circ N)$ and $(0^\circ E,0^\circ N)$. }\addu{Figure 7(a) shows the distribution of cratering rate for 3:2 resonance.
This cratering rate asymmetry has been reported in \cite{Wieczorek2012}.
They predict the cratering asymmetry maximizes at $(0^\circ E,0^\circ N)$ and $(180^\circ E,0^\circ N)$ and minimizes at $(\pm 90^\circ E,0^\circ N)$.
\delu{Because the prime meridian in this result is different from  Wieczorek {et~al.}(2012), only the relative position of maxima and minima is significant.
The distance between maxima of cratering asymmetry is $120^\circ$ in this study and $180^\circ$ in Wieczorek {et~al.}(2012).}
\addu{Both of this study and \cite{Wieczorek2012} predict the distance between maxima of cratering asymmetry is $180^\circ$.}
The difference between this study and \cite{Wieczorek2012} may be from the ignorance of the non-uniformity on the azimuth of asteroids velocity in this study \addu{and different definition of prime meridian between this study
and \cite{Wieczorek2012}}.
\addu{In Figure 7(b), the cratering with orbital eccentric $e = 0$ shows a different distribution from $e = 0.205$ for 3:2 resonance.
The longitudinal variation of cratering rate is diminished by rotation of planets and moons with eccentric $e = 0$.
However, in the cratering rate on the Moon, the difference caused by eccentric is less than 0.2\%. 
The influences of eccentric is probably related to the types of spin-orbit resonance and will be investigated in a future work.}
}

\begin{figure}
\centering\includegraphics[width=0.9\textwidth, angle=0]{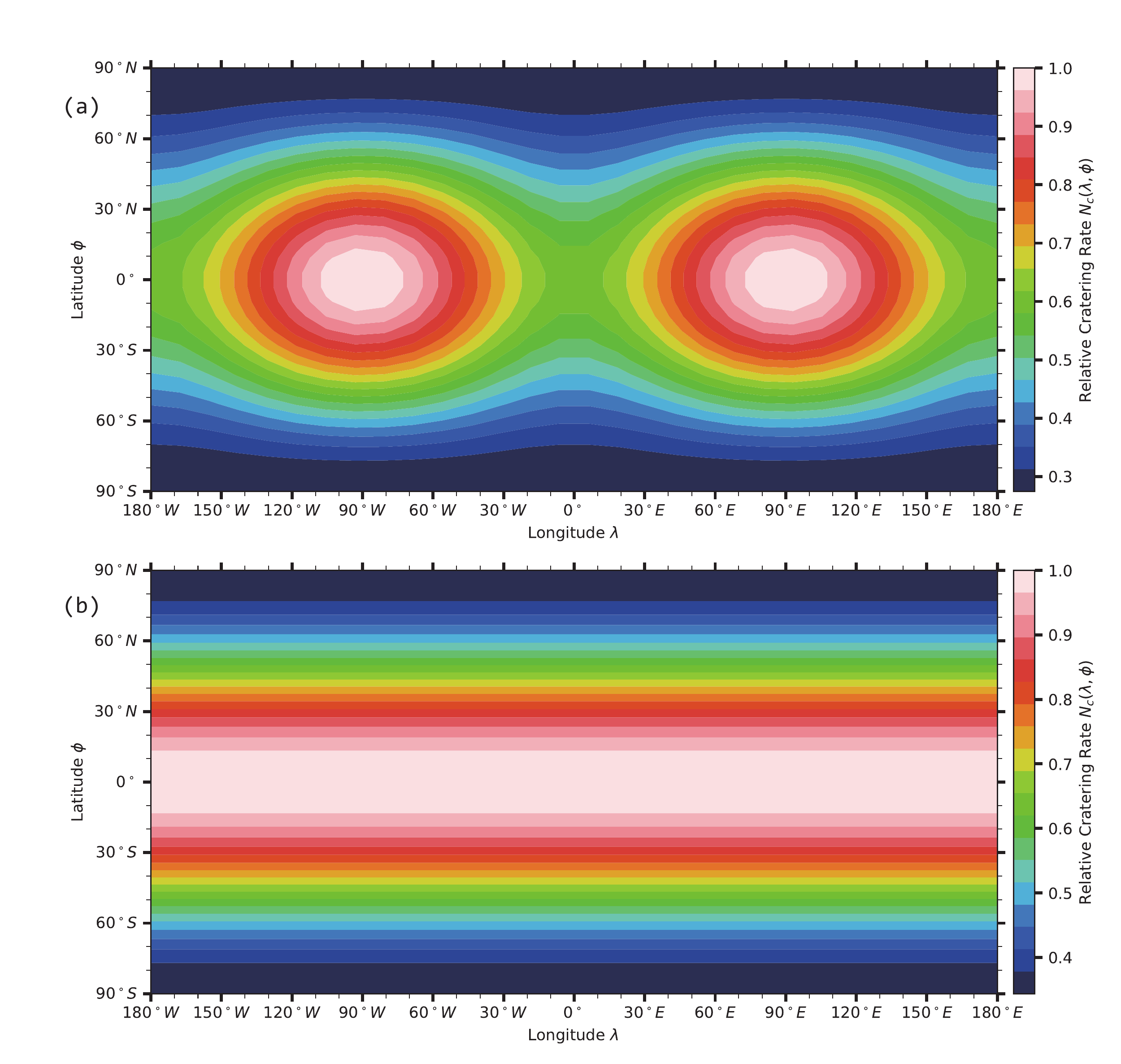}
\caption{The relative cratering rate for 3:2 resonance. The maximum is set to 1.00. longitudes $0^\circ$ and $180^\circ$ are subsolar points when $M=0$. In subfigure (a), the orbital eccentric is 0.205. In subfigure (b), the 
orbital eccentric is 0.0.}
\end{figure}

\section{Conclusion}
In this study, we have presented an extension of \cite{Wang2016} and \cite{LeFeuvre2011} to calculate the lunar cratering asymmetry with high obliquity and inclination.
Different from previous models, this model is also able to calculate the cratering asymmetry with different Cassini states, and synchronous rotating speed.
This model gives consistent results with previous with low obliquity and inclination. 
When the obliquity, inclination and Earth-Moon distance are at current values, this model gives an cratering asymmetry maximizing at $(90^\circ W,0^\circ N)$ and minimizing at \delu{$(90^\circ E,\pm 65^\circ N)$ }\addu{$(90^\circ E,\pm 53^\circ N)$} using the encountering velocity inclination distribution calculated in \cite{LeFeuvre2008}.
The apex/ant-apex ratio of this asymmetry is 1.36 and the pole/equator ratio is 0.87.
In order to calculate the cratering rate with high obliquity and inclination, we have assumed the orbital obliquity and inclination of the Moon don't affect the asteroids population encountering with the Moon. 
\addu{Increasing the} orbital obliquity and inclination of the Moon reduces the apex/ant-apex ratio.
According to the evolution of orbital obliquity and inclination of the Moon, this model gives an increasing trend in apex/ant-apex ratio with the Earth-Moon distance between $[35R_e,42R_e]$.
This ratio was predicted decreased with the increasing of Earth-Moon distance in previous studies.
Besides the cratering rate, this model also gives the spatial variation of impact flux and impact normal speed.
Our results provide the quantitative information for evaluating and rectifying the lunar cratering chronology.

\begin{acknowledgements}
    We thank M. A. Wieczorek and W. Fa for instructive discussion at the early stage of this study. 
    We also thank J.-L. Zhou for constructive and insightful suggestions on our research focus.
    We thank careful reviews by two anonymous reviewers.
    Computations were conducted on the High-performance Computing Platform of Peking University and the Pawsey Supercomputing Centre with funding from the Australian Government and the Government of Western Australia. Z.Y. is supported by the B-type Strategic Priority Program of the Chinese Academy of Sciences, Grant No. XDB41000000 and NSFC 41972321. N.Z. is grateful for NSFC 41674098, CNSA D020205 and the B-type Strategic Priority Program of the Chinese Academy of Sciences, Grant No. XDB18010104. This research has made use of data and/or services provided by the International Astronomical Union's Minor Planet Center.
\end{acknowledgements}
\bibliographystyle{raa}
\bibliography{RAA-2020-0358.R2.bib}
\appendix
\section{Asteroids encountering with the Moon with high obliquity and inclination}
In section 2.1, we \delu{thought} \addu{assumed} the concentration of asteroids encountering with the Moon is unaffected by the orbital obliquity and inclination of the Moon. 
Although the probability of asteroids encountering with the Moon has been estimated based on \citep{Opik1951,Wetherill1967,Greenberg1982} and it has been demonstrated the probability encountering with the Moon is similar with the Earth when the lunar inclination and obliquity is about 0 in Figure 6 of \cite{LeFeuvre2008}. 
But when the inclination and obliquity is high, the rationality of our assumption is uncertain. 
In this section we introduce a different framework to prove this assumption. 
For any asteroid with 
semi-major axis,
eccentricity,
inclination,
longitude of ascending node,
argument of perihelion,
true anomaly, 
mean anomaly and
eccentric anomaly are 
$(a,e,i,\Omega,\omega,f,M,E)$.
Here we use subscript $e$ to represent the orbit of Earth and $m$ to represent the Moon.
In the \addu{heliocentric} ecliptic coordinates system, the position of this asteroid is 
\begin{align}
\vec{r} &= r R_z(\Omega)R_x(i)R_z(\omega + f)[1,0,0]^T \\
r &= a(1-e^2)/(1+e\cos{f}) = a(1 -e\cos{E}) \\
& r_{max} = a(1+e) \ , \ r_{min} = a(1 -e) \nonumber 
\end{align}
For elliptic trajectory,
\begin{align}
M &= E - e\sin{E} \\
\cos{f} &= \frac{\cos{E}-e}{1-e\cos{E}} \\
\sin{f} &= \frac{\sqrt{1-e^2}\sin{E}}{1-e\cos{E}}
\end{align}
Using the common assumption: uniform precession of $\Omega$ and $\omega$\citep{Opik1951,Wetherill1967,Greenberg1982}, for given $(a,e,i)$ the
joint probability density is
\begin{align}
&P_{\Omega,\omega,M}(\Omega,\omega,M|a,e,i) = \left(\frac{1}{2\pi}\right)^3 \\
&P_{\Omega,\omega,E}(\Omega,\omega,E|a,e,i) \nonumber \\
& = P_{\Omega,\omega,M}(\Omega,\omega,M|a,e,i) |\frac{\partial M}{\partial E}|  
= \left(\frac{1}{2\pi}\right)^3 (1 + e\sin{E})\\
&\Omega \in [-\pi,\pi],\ \omega \in [-\pi,\pi],\ M \in [-\pi,\pi],
\ E \in [-\pi,\pi]
\end{align}
The position of this asteroid can also be expressed as $\vec{r}=[x,y,z]^T$, then we obtain a transformation: $(\Omega,\omega,E) \mapsto (x,y,z)$.
\begin{align}
    \left\{\begin{array}{cll}
    x &= & r (\cos{\Omega}\cos{(\omega + f)} - \sin{\Omega}\sin{(\omega + f)}\cos{i}) \\
    y &= & r (\sin{\Omega}\cos{(\omega + f)} + \cos{\Omega}\sin{(\omega + f)}\cos{i}) \\
    z &= & r \sin{(\omega + f)}\sin{i}
    \end{array}\right.
\end{align}
Eq. (A.9) has 4 solutions: $(\Omega_k,\omega_k,E_k), k = 1,2,3,4$.
\begin{align}
    \left\{\begin{array}{cll}
    \cos{f_k} &= & 
        \cfrac{a(1-e^2)}{e\sqrt{x^2 + y^2 + z^2}}-\cfrac{1}{e} \\
    \sin{(\omega_k + f_k)} & = & 
        \cfrac{z}{\sqrt{x^2 + y^2 + z^2}\sin{i}}
    \\
    \Omega_k &= & atan2\Big(x\cos{(\omega_k+f_k)} + y\sin{(\omega_k+f_k)}\cos{i}, \\
    & & y\cos{(\omega_k+f_k)} - x\sin{(\omega_k+f_k)}\cos{i}\Big) \\
    E_k &=& atan2\Big(
        \sqrt{x^2 + y^2 + z^2}\cos{f_k} + ae, \nonumber \\
        & & \sqrt{x^2 + y^2 + z^2}\sin{f_k}\frac{1}{\sqrt{1-e^2}} \Big) 
    \end{array}\right. \nonumber \\
\end{align}
Using Eqs. (A.6-A.10), the joint probability density is
\begin{align}
    & P_{x,y,z}(x,y,z|a,e,i) = \sum_{k=1}^{4} 
    \left| \left|
    \cfrac{\partial(\Omega_k,\omega_k,E_k)}{\partial(x,y,z)}
    \right| \right|
    P_{\Omega,\omega,E}(\Omega_k,\omega_k,E_k|a,e,i) \nonumber \\
    & = \frac{1}{2a\pi^3}\frac{1}{\sqrt{r^2 \sin^2{i} -z^2}} \frac{1}{\sqrt{(r-r_{min})(r_{max}-r)}} \ , \  \ r = \sqrt{x^2 + y^2 + z^2}
\end{align}

In Eq. (A.11), $\cfrac{\partial(\Omega_k,\omega_k,E_k)}{\partial(x,y,z)}$ is the Jacobi matrix. Eq. (A.11) is valid when $(x,y,z) \in \mathbf{D} = \{
|z| \le |\sin{i}| r \ and \ 
r_{min} \le r \le r_{max}\}$. When $(x,y,z) \notin \mathbf{D}$, $P_{x,y,z}(x,y,z|a,e,i) = 0$.  This asteroid encountering with a fix point $\vec{r_0} = [x_0,y_0,z_0]^T$ is defined as $|\vec{r} - \vec{r_0}| \le \tau$ ($\tau$ is different for the Moon and the Earth) and $\tau \ll min\{|\vec{r}|,|\vec{r_0}|\} $. Then we obtain the probability encountering with a fixed point $P_1$ and its error $\delta P_1$.
\begin{align}
    P_1 &= \iiint_{|\vec{r} - \vec{r_0}| \le \tau} P_{x,y,z}(x,y,z|a,e,i)
    dxdydz \nonumber \\
    & \approx \frac{4}{3}\pi\tau^3 P_{x,y,z}(x_0,y_0,z_0|a,e,i) \\
   \delta P_1 
    & \approx \iiint_{|\vec{r} - \vec{r_0}| \le \tau} 
    |\vec{r} - \vec{r_0}| |\nabla P_{x,y,z}(x_0,y_0,z_0|a,e,i)|
    dxdydz \nonumber \\
    & = \pi \tau^4 |\nabla P_{x,y,z}(x_0,y_0,z_0|a,e,i)|
\end{align}
Because $P_{x,y,z}$ is not bounded with $(r^2\sin^2{i} - z^2)(r-r_{max})(r-r_{min}) = 0$. Eq.(A.12) and Eq. (A.13) are valid when $min\{|r\sin{i} \pm z|,|r-r_{max}|,|r-r_{min}| \} \ge \varepsilon a \ (\varepsilon>0)$.
When $min\{|r\sin{i} \pm z|,|r-r_{max}|,|r-r_{min}| \} < \varepsilon a$, the supremum of $P_1$ can be estimated in spherical coordinates,
\begin{align} 
& P_1 \le \iiint_{E} P_{x,y,z} r^2 |\sin{\theta}| dr d\theta d\varphi 
= \frac{1}{2 a \pi^3}
\left(\int_{\varphi} d\varphi\right) \nonumber \\
& \cdot Re\left(\int_{\theta} \frac{\sin{\theta} d\theta}{\sqrt{\sin^2{i} - \cos^2{\theta}}}\right)
Re\left(\int_{r} \frac{r dr}{\sqrt{(r-r_{min})(r_{max}-r)}} \right)
\\
& E = \Big\{
|r-r_0|\le \tau, 
|\theta -\theta_0| \le \arctan{\tau/r_0}, 
|\varphi -\varphi_0| \le 
\arctan{\frac{\tau}{r_0 \cos{\theta_0}}}
\Big\}
\end{align} 
This section is only a qualitative explanation. For simplicity, following derivations are under the condition:
 $min\{|r\sin{i} \pm z|,|r-r_{max}|,|r-r_{min}| \} \ge \varepsilon a$.
When $\vec{r_0}$ is not fixed, for the Earth $\vec{r_0} = \vec{r_e}
= r_e R_z(\Omega_e)R_x(i_e)R_z(\omega_e + f_e)[1,0,0]^T$, this asteroid
encounters with the Earth by probability $P_2$. 
\begin{align}
    & P_2 = \iiint_{(\Omega_e,\omega_e,M_e)} P(\Omega_e,\omega_e,M_e|a_e,e_e,i_e) P_1 d\Omega_e d\omega_e dM_e \nonumber \\
    & = 2\pi \iint_{(\omega_e,M_e)} P(\Omega_e = 0,\omega_e,M_e|a_e,e_e,i_e) P_1 d\omega_e dM_e \\
    & \delta P_2 = 2\pi\iint_{(\omega_e,M_e)} P(\Omega_e = 0,\omega_e,M_e|a_e,e_e,i_e) \delta P_1 d\omega_e dM_e
\end{align}
Eq. (A.16) and Eq. (A.17) use the rotational symmetry about the z axis of $P_{x,y,z}$.
For the Moon 
$\vec{r_0} = \vec{r_e} + \vec{r_m}=r_e R_z(\Omega_e)R_x(i_e)R_z(\omega_e + f_e)[1,0,0]^T + r_m R_z(\Omega_m)R_x(i_m)R_z(\omega_m + f_m)[1,0,0]^T$
, this asteroid encounters with the Moon by $P_3$.
\begin{align}
    &P_3 = 2\pi \iint_{(\omega_e,M_e)} P(\Omega_e=0,\omega_e,M_e|a_e,e_e,i_e) \nonumber \\
    & \iiint_{(\Omega_m,\omega_m,M_m)}  P(\Omega_m,\omega_m,M_m|a_m,e_m,i_m) P_1  d\omega_e dM_e d\Omega_m d\omega_m dM_m \\
    &\delta P_3 = 2\pi \iint_{(\omega_e,M_e)} P(\Omega_e=0,\omega_e,M_e|a_e,e_e,i_e) \nonumber \\
    &\iiint_{(\Omega_m,\omega_m,M_m)}  P(\Omega_m,\omega_m,M_m|a_m,e_m,i_m) \delta P_1  d\omega_e dM_e d\Omega_m d\omega_m dM_m 
\end{align}
The difference between $P_2$ and $P_3$ can be estimated by 
\begin{align}
&|P_2\tau_e^{-3} - P_3\tau_m^{-3}| \le 2\pi \iint_{(\omega_e,M_e)} P(\Omega_e=0,\omega_e,M_e|a_e,e_e,i_e) 
\iiint_{(\Omega_m,\omega_m,M_m)}
P(\Omega_m,\omega_m,M_m|a_m,e_m,i_m) 
\nonumber \\  
&  \times \frac{4\pi}{3} | P_{x,y,z}(x_m+x_e,y_m+y_e,z_m+z_e|a,e,i) 
 - P_{x,y,z}(x_e,y_e,z_e|a,e,i) | d\omega_e dM_e d\Omega_m d\omega_m dM_m
\end{align}
From Eq. (A.20), $P_3$ can be estimated by $P_2 \frac{\tau_m^3}{\tau_e^3}$. While $P_2 \frac{\tau_m^3}{\tau_e^3}$ is independent of lunar inclination and obliquity, therefore we can use the concentration of asteroids encountering with the Moon with low inclination and obliquity 
to replace the concentration with high inclination and obliquity. 
We note that when $(a_e,i_e,e_e) = (1,0,0)$, for $87\%$ of the near-earth orbits (the dataset of near-earth orbts is taken from the International Astronomical Union’s website), the relative error between $P_2 \frac{\tau_m^3}{\tau_e^3}$ and $P_3$ calculated by Eq. (A.20) is less than 5\%.
\end{document}